\documentclass[journal,twoside,web]{ieeecolor}
\usepackage{tmi}
\usepackage{cite}
\usepackage{amsmath,amssymb,amsfonts}
\usepackage{algorithm}
\usepackage{algorithmic}
\usepackage{graphicx}
\usepackage{textcomp}

\def\z{{\mathbf z}}

\def\1{{\mathbf 1}}

\def\X{{\mathbf X}}

\def\BibTeX{{\rm B\kern-.05em{\sc i\kern-.025em b}\kern-.08em
    T\kern-.1667em\lower.7ex\hbox{E}\kern-.125emX}}

\begin{document}

\title{An Adaptive, Disentangled Representation for Multidimensional MRI Reconstruction}

\author{Ruiyang Zhao and Fan Lam
\thanks{
This work was supported in part by the following grants: NSF-CBET 1944249, NIH/NIGMS 5R35GM14296.  }
\thanks{R. Zhao is with the Department of Electrical and Computer Engineering and Beckman Institute for Advanced Science and Technology, University of Illinois Urbana-Champaign.\\
\indent F. Lam is with the Department of Bioengineering and Beckman Institute for Advanced Science and Technology, University of Illinois Urbana-Champaign, IL 61801 USA (e-mail: fanlam1@illinois.edu).}}
\maketitle

\begin{abstract}
We present a new approach for representing and reconstructing multidimensional magnetic resonance imaging (MRI) data. Our method builds on a novel, learned feature-based image representation that disentangles different types of features, such as geometry and contrast, into distinct low-dimensional latent spaces, enabling better exploitation of feature correlations in multidimensional images and incorporation of pre-learned priors specific to different feature types for reconstruction. More specifically, the disentanglement was achieved via an encoder-decoder network and image transfer training using large public data, enhanced by a style-based decoder design. A latent diffusion model was introduced to impose stronger constraints on distinct feature spaces. New reconstruction formulations and algorithms were developed to integrate the learned representation with a zero-shot self-supervised learning adaptation and subspace modeling. The proposed method has been evaluated on accelerated $\text{T}_1$ and $\text{T}_2$ parameter mapping, achieving improved performance over state-of-the-art reconstruction methods, without task-specific supervised training or fine-tuning. This work offers a new strategy for learning-based multidimensional image reconstruction where only limited data are available for problem-specific or task-specific training.
\end{abstract}


\section{Introduction}
The success of deep learning has facilitated the paradigm shift in image reconstruction from generic ``hand-crafted'' regularization/priors to learning and incorporating domain/task-specific priors via data-driven methods. In the realm of MRI, one mainstream approach is unrolling iterative optimization algorithms into cascaded deep neural networks that map directly from an initial reconstruction (or $k$-space measurements) to the reconstructed image. These networks are trained end-to-end in a supervised fashion when high-quality, fully-sampled images are available \cite{hammernik2018learning,sun2016deep,MoDL,monga2021algorithm}, or in a self-supervised manner leveraging only noisy or incomplete measurements \cite{yaman2020self,akccakaya2022unsupervised,millard2023theoretical}. While both training strategies have demonstrated impressive performance, they typically require large quantities of training data to achieve reliable results. Furthermore, the task-specific nature of these approaches renders the learned models vulnerable to variations in acquisition protocols, data quality, and domain shifts \cite{darestani2021measuring,heckel2024deep}.

An alternative learning-based image reconstruction approach is to pre-learn a ``prior'' that can be flexibly incorporated (or ``plugged in'') into iterative algorithms considering task-specific forward models, rather than learning the inverse mapping end-to-end. For example, prior learning techniques such as score-based diffusion models have been extensively investigated \cite{vincent2011connection, ho2020denoising, song2019generative}, along with various algorithms designed to integrate these learned priors\cite{gungor2023adaptive,kamilov2023plug,chung2022diffusion,webber2024diffusion}. These methods circumvent the need of retraining for each new reconstruction task and can be more robust to acquisition and domain variations. While promising, their application in multidimensional imaging remains largely unexplored due to the scarcity of training data and challenges in effectively leveraging feature correlations across multiple image dimensions. For example, in multidimensional MRI applications such as quantitative parameter mapping \cite{ma2013magnetic,ma2020three,meng2021accelerating}, MR spectroscopic imaging \cite{posse2013mr,lam2023high_SPM}, or dynamic imaging \cite{oscanoa2023deep,christodoulou2020accelerated}, acquiring large and diverse datasets for learning the entire spatiotemporal prior is often prohibitive considering the resources needed. Even with adequate data, strategies to more effectively exploit multidimensional feature correlations are needed.

Another approach that has attracted more attention recently is using network-based representation of unknown images for zero-shot learning-based reconstruction. In this approach, the target image is modeled as the output of a network, typically via a decoder architecture that maps a low-dimensional feature vector (latent) into images. Compared to other learning-based methods, it offers several advantages: (1) it substantially reduces the degrees of freedom by reformulating image reconstruction as the recovery of low-dimensional latents; (2) it enhances generalizability across tasks, since the network learns image representations rather than direct inverse mapping; (3) it offers flexibility in data-limited settings, as the representation can be implemented using untrained networks (e.g., deep image prior \cite{ulyanov2018deep,yoo2021time}), pre-trained networks (e.g., generative adversarial networks\cite{bora2017compressed} and latent diffusion models\cite{song2023solving}), or pre-trained networks plus experiment/task-specific fine-tuning or adaptation \cite{hussein2020image,narnhofer2019inverse}; and (4) by describing images via a set of features rather than individual voxel values, it enables new ways to constrain multidimensional images that often exhibit correlated features. For example, in quantitative MRI (qMRI), image contrast changes with acquisition parameters while the underlying anatomy (geometry) remains unchanged. Therefore, network-based representation offers a potential solution to the data scarcity and generalization challenges, particularly in multidimensional acquisition settings.

In this work, we proposed a new network-based representation for reconstructing multidimensional MR images. Specifically, we introduced a model and learning strategy that separate different types of image features such as geometry and contrast into distinct low-dimensional latent spaces, reducing the degrees of freedom for multidimensional images and enabling flexible constraints on individual features. Latent diffusion models were employed to provide feature-level generative priors  on the disentangled latents for constrained reconstruction. To mitigate potential mismatch between representations pre-trained using large public datasets and application-specific data, we developed a novel algorithm that combines pretrained representations with task-specific adaptation through zero-shot self-supervised learning. We demonstrated effective disentanglement achieved for geometry and contrast features in multi-contrast images and evaluated its utility in multidimensional MR applications such as accelerated $\text{T}_1$ and $\text{T}_2$ mapping.

The remaining of the paper is organized as follows: Section
II provides some background on feature-base image representation and its use in image reconstruction. Section III describes the proposed problem formulation and algorithm in details. Section IV evaluates the proposed method in two application examples, accelerated $\text{T}_1$ and $\text{T}_2$ mapping. Section V and VI provide some
technical discussion and conclude the paper.

\section{Background}
\subsection{Feature-based Representation and Reconstruction}
In feature-based representation, the image of interest $\mathbf{X}$ is described as: $\mathbf{X} = D_{\boldsymbol{\theta}}(\mathbf{z})$, with $\mathbf{z}$ denoting a set of latent variables (typically lower dimensional than $\X$) and $D_{\boldsymbol{\theta}}(.)$ a network parameterized by $\boldsymbol{\theta}$ mapping the latent to image space. Compared to the earlier kernel-based representation \cite{wang2014pet,li2018constrained}, these network-based models offer more flexibility and circumvent the need to choose a specific kernel. With this model, the reconstruction problem can be formulated as estimating $\mathbf{z}$ and/or the network parameters $\boldsymbol{\theta}$ by solving:
\begin{equation}
\mathbf{\hat{z}},\boldsymbol{\hat{\theta}}=\arg\min _{\mathbf{z},\boldsymbol{\theta}} \left\|\mathbf{A}(D_{\boldsymbol{\theta}}(\mathbf{z}))-\mathbf{y}\right\|_2^2, 
\label{Background_recon}
\end{equation}
where $\mathbf{A}$ and $\mathbf{y}$ denote the forward model and measured data, respectively. Since the reconstruction is constrained to lie within the range space of $D_{\boldsymbol{\theta}}(\cdot)$, this approach may introduce substantial modeling errors from several sources: mismatch between the training data for representation learning and test data, low-dimensionality of $\z$, and limitations of optimization algorithms. This issue has been observed in methods based on pretrained GANs that update only the latent variables $\mathbf{z}$ \cite{bora2017compressed}. While jointly updating both $\mathbf{z}$ and $\boldsymbol{\theta}$ can mitigate this issue, it requires early stopping and/or other careful regularizations to prevent overfitting to noise and artifacts. 

To reduce representation error while maintaining effective constraints, we will introduce a second-stage refinement network combined with a zero-shot self-supervised learning strategy to update the network parameters and latent-diffusion-based constraints. In addition, the proposed network is adapted from a style-based architecture \cite{karras2019style,karras2020analyzing}, which has demonstrated stronger image representation capability than earlier generations of GANs, particularly for high-resolution images.

\begin{figure}[!t]
\centerline{\includegraphics[width=0.9\columnwidth]{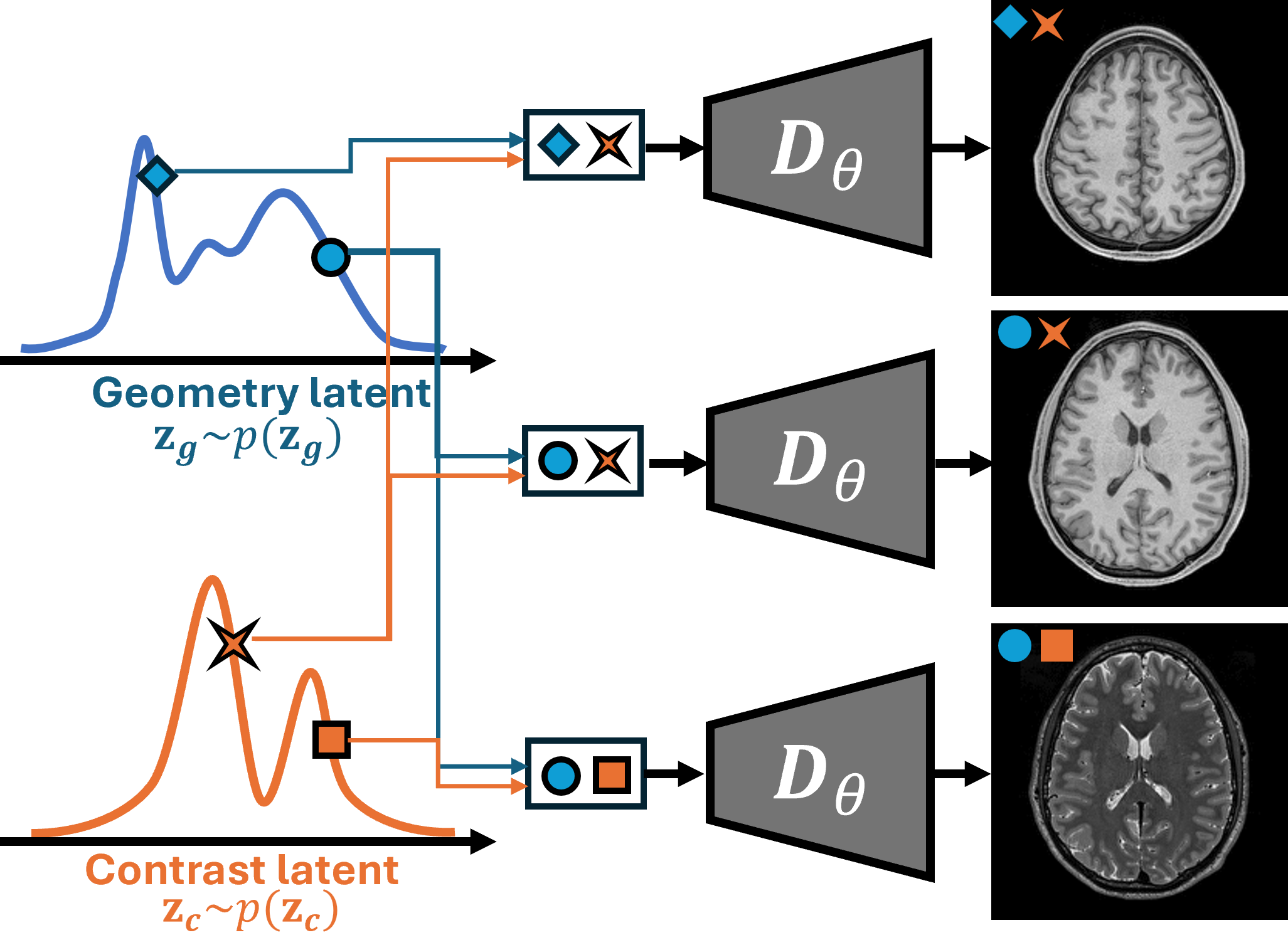}}
\caption{A disentangled representation (using “geometry” and “contrast” features as examples): Once learned, geometry and contrast latents can be sampled from respective distributions and combined to generate images with target geometry or contrast.}
\label{Figure_disentangle_rep}
\end{figure}

\subsection{Multidimensional Image Representation}
A straightforward extension of feature-based image representation to multidimensional imaging is to assign separate latents and network parameters to each image, i.e.,
\begin{equation}
\mathbf{X}_1 = D_{\boldsymbol{\theta}_1}(\mathbf{z}_1),\ \mathbf{X}_2 = D_{\boldsymbol{\theta}_2}(\mathbf{z}_2),\ \dots,\ \mathbf{X}_{N_t} = D_{\boldsymbol{\theta}_{N_t}}(\mathbf{z}_{N_t}),
\end{equation}
where $\{\mathbf{X}_1, \mathbf{X}_2, \dots, \mathbf{X}_{N_t}\}$ denote the images acquired at different times or parameters. However, modeling each image independently ignores their inherent correlations and introduces unnecessarily more unknowns. To this end, a common strategy is to enforce shared network parameters across all images ($\boldsymbol{\theta}_1=\boldsymbol{\theta}_2=\dots=\boldsymbol{\theta}_{N_t}$) and explore correlations in latent space. For example,  similarity constraints can be introduced on adjacent latents (e.g., $\mathbf{z}_i \approx \mathbf{z}_{i+1}$) by assuming that images acquired at neighboring time points share similar features. Additionally, multi-resolution architectures like StyleGAN can be utilized to impose invariance on certain latent components across images \cite{karras2019style,karras2020analyzing,kelkar2021prior}. Although helpful, these methods rely on generic similarity assumptions that do not capture the true relationships among the latents for multidimensional images.

To better model feature correlations in the latent space, we propose learning a disentangled representation that decomposes the latent space into semantically distinct components. Considering contrast and geometry variations as an example, images with different contrast weightings $\{\mathbf{X}_{c_1}, \mathbf{X}_{c_2}, \ldots, \mathbf{X}_{c_{N_t}}\}$ are acquired in many applications such as quantitative MRI. If contrast variations can be modeled separately from geometry in the latent space (see Fig.~\ref{Figure_disentangle_rep}), each image can be modeled using a shared geometry latent $\mathbf{z}_g$ and contrast-specific latents $\{\mathbf{z}_{c_1}, \mathbf{z}_{c_2}, \ldots, \mathbf{z}_{c_{N_t}}\}$, which leads to: $\mathbf{X}_{c_i} = D_{\boldsymbol{\theta}}(\mathbf{z}_g, \mathbf{z}_{c_i})$ for $i = 1, 2, \ldots, N_t$. This disentangled representation better exploits feature correlations and enables explicit control over distinct image features. The next section describes our detailed methodology.

\section{Proposed Method}
\subsection{Learning A Disentangled Image Representation}
\begin{figure}[!t]
\centerline{\includegraphics[width=\columnwidth]{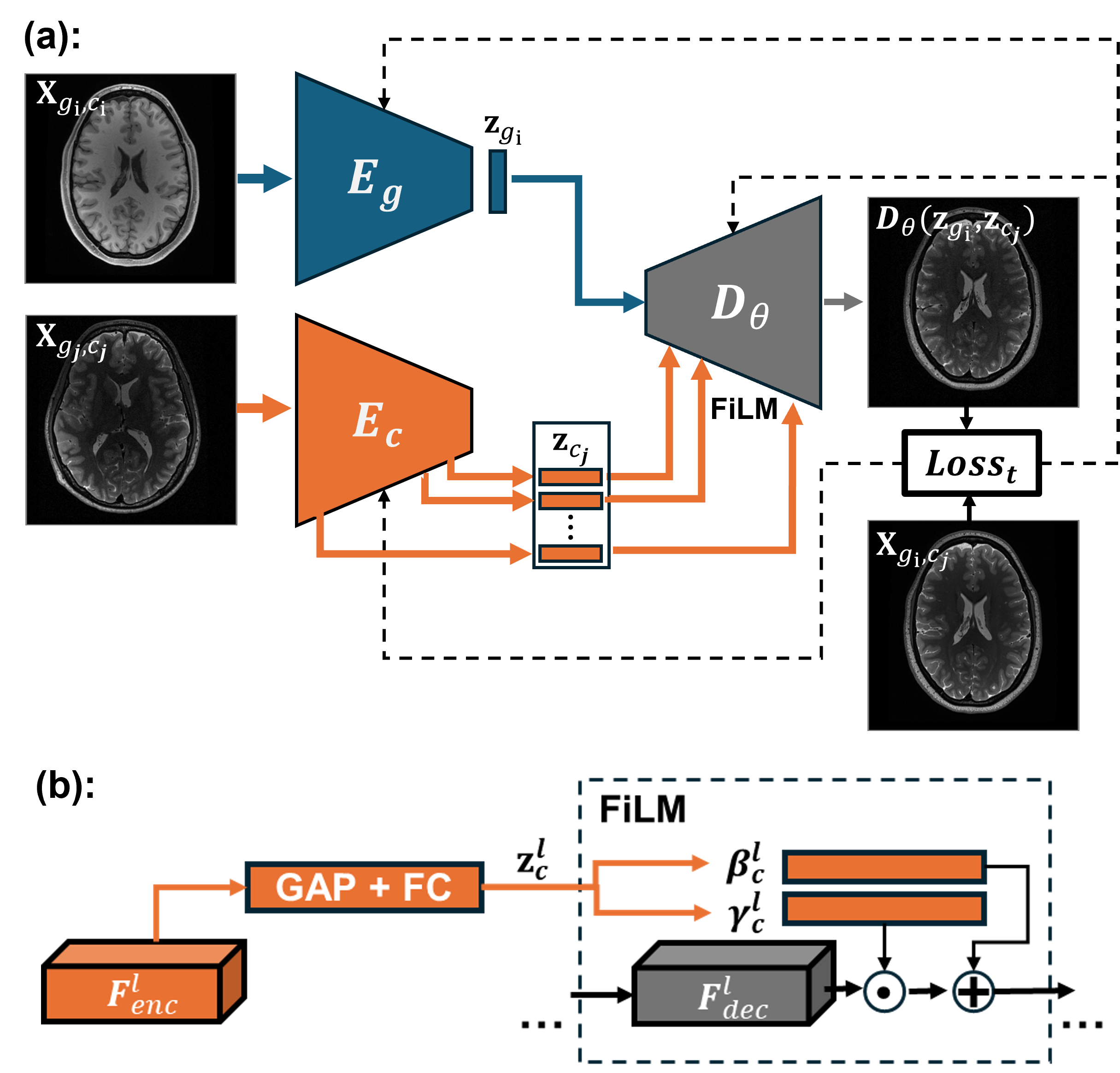}}
\caption{(a) The proposed representation learning strategy. Image transfer losses are used to train the disentangled representation. The decoder combines geometry latents from $\mathbf{X}_{g_i,c_i}$ with contrast latents from $\mathbf{X}_{g_j,c_j}$ to generate a new image (see right), which can be interpreted as contrast transfer for $\mathbf{X}_{g_i,c_i}$ or geometry transfer for $\mathbf{X}_{g_j,c_j}$. (b) The FiLM block for feature combination. At resolution $l$, the encoder feature maps $\mathbf{F}_{enc}^l$ are transformed into $\mathbf{z}_c^l$ through global average pooling (GAP) and a fully connected (FC) layer. The resulting $\mathbf{z}_c^l$ is then split into modulation parameters $\boldsymbol{\gamma}_c^{l}$ and $\boldsymbol{\beta}c^{l}$, which are applied to modulate the decoder feature maps $\mathbf{F}_{dec}^l$ at the same level.}
\label{fig:Figure_representation learning}
\end{figure}
Our target is to learn a model that enables explicit control of different features in the target images. To this end, we propose a two-step learning strategy. The first step involves training an autoencoder to enable feature disentanglement through image transfer and latents regularization, while the second step focuses on learning priors distributions for disentangled latents using diffusion models. In this paper, we investigate the disentanglement of contrast and geometry features, with a specific focus on reconstructing images with different contrast-weightings, while the approach can be generalized to other types of features.

More specifically, we first represent multicontrast images using two distinct sets of latents, i.e., geometry and contrast. Two separate image encoders, $E_g(\cdot)$ and $E_c(\cdot)$, were employed to extract geometry and contrast information, respectively. The resulting latents were then combined by a decoder $D_{\boldsymbol{\theta}}(\cdot)$ for image synthesis (Fig.~\ref{fig:Figure_representation learning}). The autoencoder training is based on image transfer, where the decoder combines a geometry latent from one image with a contrast latent from another to synthesize cross-composed images, minimizing an image transfer loss:
\begin{equation}
    \mathcal{L}_{\text{t}} = \left\| \mathbf{X}_{g_i,c_j} - D_{\boldsymbol{\theta}}(E_g(\mathbf{X}_{g_i,c_i}), E_c(\mathbf{X}_{g_j,c_j})) \right\|_F^2, \label{loss_cst} 
\end{equation}
where $\mathbf{X}_{g_i,c_{i}}$ and $\mathbf{X}_{g_{j},c_j}$ denotes image pairs with geometry $g_i$ and $g_j$ and contrast $c_i$ and $c_j$. The target $\mathbf{X}_{g_i,c_j}$ corresponds to an image sharing the same geometry as $\mathbf{X}_{g_{i},c_i}$ and the same contrast as $\mathbf{X}_{g_j,c_j}$. 

\begin{figure}[!t]
\centerline{\includegraphics[width=\columnwidth]{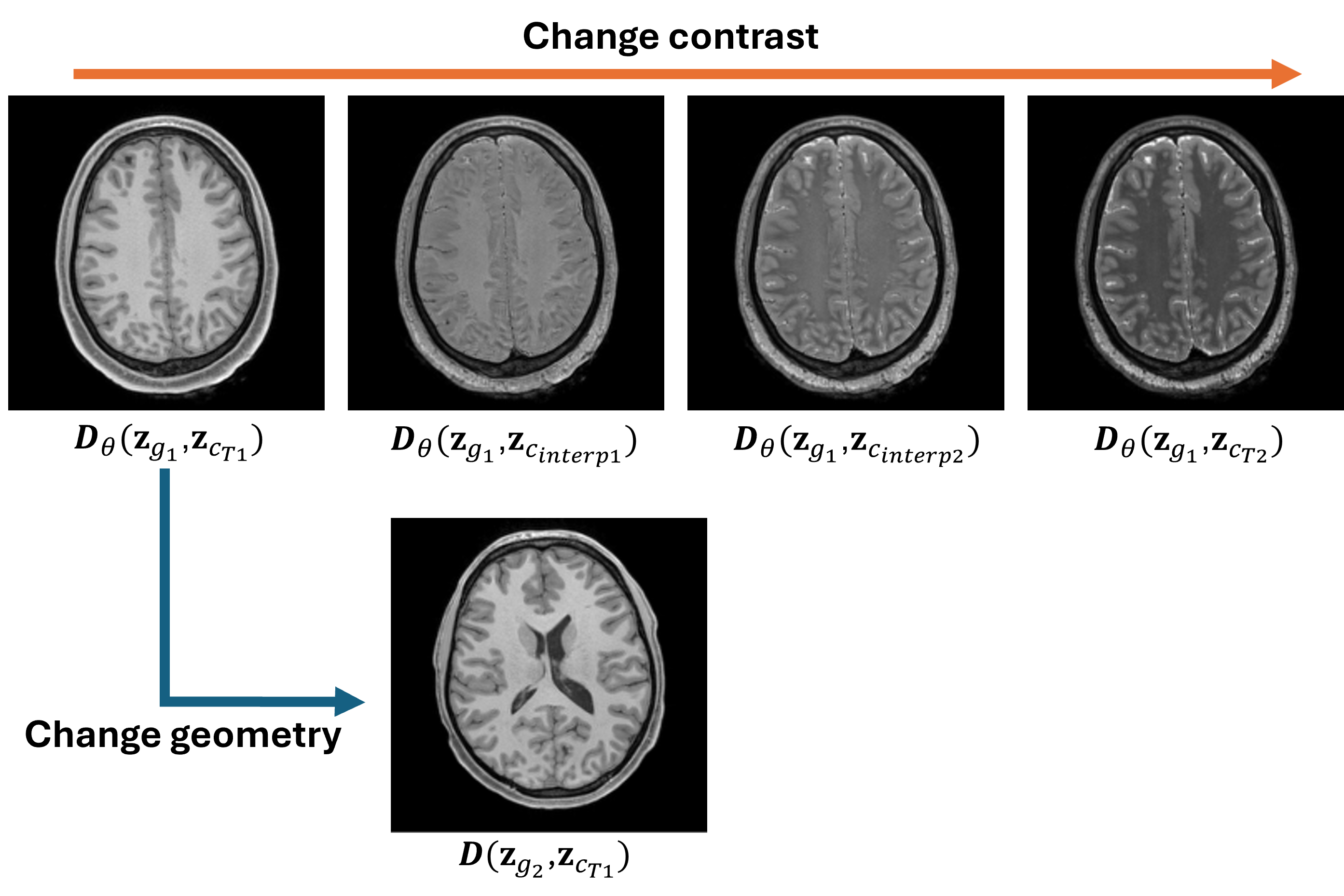}}
\caption{Feature disentanglement achieved: High-quality $\text{T}_1$w (first) and $\text{T}_2$w (last) images can be generated by sampling a geometry latent $\mathbf{z}_{g_{1}}$ from our learned latent diffusion model and combining it with two different contrast latents $\mathbf{z}_{c_{T_1}}$ and $\mathbf{z}_{c_{T_2}}$, using the trained decoder. Interpolation between $\mathbf{z}_{c_{T_1}}$ and $\mathbf{z}_{c_{T_2}}$ produced images with consistent geometry but varying contrasts (middle images). Combining a different geometry latent with $\mathbf{z}_{c_{T_1}}$ produced the same $\text{T}_1$ contrast but with different anatomical features (bottom).}
\label{disentangle_results}
\end{figure}

\begin{algorithm*}[t]
\caption{The Proposed Reconstruction Algorithm with the Learned Representation and Latent Diffusion Prior}
\label{alg:resample}
\begin{algorithmic}[1]
\REQUIRE Measurements $\mathbf{d}$, Forward Operator $\mathbf{A}(\cdot)$, Decoder $D_{\boldsymbol{\theta}}(\cdot)$, Refinement Network $N_{\boldsymbol{\theta}_N}(\cdot)$, Pretrained Noise Predictor $\epsilon_{\boldsymbol{\theta}_{g}}(\cdot)$, Diffusion Parameters $\bar{\alpha}_{t_d}, \eta, \delta$, Hyperparameter $\gamma$, Data Consistency Update Steps $C$ 
\STATE Initialize $\{\mathbf{z}_{g_t,T_d}\}, \{{\mathbf{z}}_{c_t}\} \sim \mathcal{N}(0,I)$ and refinement params $\theta_N$
\FOR{$t_d = T_d-1, \ldots, 0$}
  \STATE \textbf{Diffusion step}: Predict noise $\{\hat{\epsilon}_{t,t_d+1}\} = \epsilon_{\boldsymbol{\theta}_{g}}(\{\mathbf{z}_{g_t,t_d+1}\}, t_d+1)$
  \STATE Estimate clean latent via Tweedie’s formula: $\{\bar{\mathbf{z}}_{g_t,0}\} = \tfrac{1}{\sqrt{\bar{\alpha}_{t_d+1}}}\Big(\{\mathbf{z}_{g_t,t_d+1}\} - \sqrt{1-\bar{\alpha}_{t_d+1}} \{\hat{\epsilon}_{t,t_d+1}\}\Big)$
  \STATE Update via DDIM:
  $ \{\mathbf{z}'_{g_t,t_d}\} = \sqrt{\bar{\alpha}_{t_d}}\{\bar{\mathbf{z}}_{g_t,0}\} + \sqrt{1-\bar{\alpha}_{t_d}-\eta\delta^2}\{\hat{\epsilon}_{t,t_d+1}\} + \eta\delta \{\epsilon\}$, where $\epsilon \sim \mathcal{N}(0,I)$
  \IF{$t \in C$ (data consistency steps)} \STATE \textbf{Data consistency update:} 
    \[
    \{\hat{\mathbf{z}}_{g_t}\},\{\hat{\mathbf{z}}_{c_t}\},\hat{\boldsymbol{\theta}}_N\;=\;\arg\min_{\{\mathbf{z}_{g_t}\},\{\mathbf{z}_{c_t}\},{\boldsymbol{\theta}}_N}\;
    \underbrace{\|\mathbf{A}_\Theta(\{D_{\boldsymbol{\theta}}(\mathbf{z}_{g_t},\mathbf{z}_{c_t})\})-\mathbf{d}_\Theta\|_F^2}_{\text{measured part}}
    +\underbrace{\|\mathbf{A}_\Lambda(N_{\boldsymbol{\theta}_N}(\{D_{\boldsymbol{\theta}}(\mathbf{z}_{g_t},\mathbf{z}_{c_t})\},\mathbf{d}_\Theta,\mathbf{A}_\Theta))-\mathbf{d}_\Lambda\|_F^2}_{\text{held-out part}}
    \]
    \STATE Resample(blend DDIM and DC): $\{\mathbf{z}_{g_t,t_d}\}= \text{StochasticResample}(\{\hat{\mathbf{z}}_{g_t}\}, \{\mathbf{z}'_{g_t,t_d}\}, \gamma)$
  \ELSE
    \STATE $\{\mathbf{z}_{g_t,t_d}\} = \{\mathbf{z}'_{g_t,t_d}\}$
  \ENDIF
\ENDFOR
\STATE \textbf{Final reconstruction}: $\mathbf{X} = N_{\hat{\boldsymbol{\theta}}_N}(\{D_{\boldsymbol{\theta}}(\hat{\mathbf{z}}_{g_t},\hat{\mathbf{z}}_{c_t})\},\mathbf{d},\mathbf{A})$
\end{algorithmic}
\end{algorithm*}

To further promote disentanglement, we introduced a multi-resolution, StyleGAN-like architecture. 
The contrast encoder $E_c(\cdot)$ produces latents at multiple resolutions, which are injected into 
the decoder via Feature-wise Linear Modulation (FiLM)~\cite{perez2018film}. For example, as shown in Fig.~\ref{fig:Figure_representation learning}(b), the feature map $\mathbf{F}_{\mathrm{enc}}^{l}$ extracted by $E_c(\cdot)$ at a given resolution level $l$ is transformed into 
scale and shift vectors $\boldsymbol{\gamma}^{l}, \boldsymbol{\beta}^{l} \in \mathbb{R}^{B \times C_{l}}$, 
where $B$ denotes the batch size and $C_{l}$ the number of channels. The corresponding decoder 
features $\mathbf{F}_{\mathrm{dec}}^{l} \in \mathbb{R}^{B \times C_{l} \times H_{l} \times W_{l}}$ 
are modulated as:
\begin{equation}
\operatorname{FiLM}\!\left(\mathbf{F}_{\mathrm{dec}}^{l} \,\middle|\, 
\boldsymbol{\gamma}^{l}, \boldsymbol{\beta}^{l}\right)
= \boldsymbol{\gamma}^{l} \odot \mathbf{F}_{\mathrm{dec}}^{l} + \boldsymbol{\beta}^{l},
\end{equation}
where $\odot$ is element-wise multiplication, with 
$\boldsymbol{\gamma}^{l}$ and $\boldsymbol{\beta}^{l}$ broadcasted over the spatial dimensions $(H_{l}, W_{l})$. This mechanism encourages the model to represent contrast as global, layer-wise intensity transformations while preserving spatial structure in the decoder. By explicitly modulating feature at multiple resolutions, FiLM offers an effective way to control contrast-related variations without altering spatial content, thereby promoting geometry–contrast disentanglement.

To model the distributions of latents $\z_g$ (geometry) and $\z_c$ (contrast), we trained a latent diffusion model that learns “disentangled” generative priors for multicontrast images. The loss function for training the diffusion model is defined as:
\begin{equation}
    \mathcal{L}_{\text{LDM}_{g/c}} = \mathbb{E}_{\mathbf{z}_{g/c}, \epsilon \sim \mathcal{N}(0,1), t} \left[ 
        \left\| \epsilon - \epsilon_{\boldsymbol{\theta}_{g/c}} \left( \mathbf{z}_{g/c}, t \right) \right\|_F^2 
    \right]. \label{loss_ldm}
\end{equation}
where $t$ is the time step in diffusion models, $\boldsymbol{\theta}_g$ and $\boldsymbol{\theta}_c$ denotes the parameters for the score functions of $\z_g$  and $\z_c$, respectively. We used a time-conditioned UNet as the backbone of $\epsilon_{\boldsymbol{\theta}_{g/c}}$~\cite{rombach2022high}. The model was trained using large-scale public datasets including HCP and Kirby21 datasets~\cite{van2013wu,landman2011multi}.

\subsection{Reconstruction Formulation and Algorithm}
Leveraging the pre-trained representation $D_{\boldsymbol{\theta}}(.)$, we propose the following multidimensional reconstruction formulation:
\begin{equation}
\begin{split}
&\{\hat{\mathbf{z}}_{g_t}\},\{\hat{\mathbf{z}}_{c_t}\},\hat{\boldsymbol{\theta}}_N = 
\arg \underset{\{\mathbf{z}_{g_t}\},\{\mathbf{z}_{c_t}\},\boldsymbol{\theta}_N}{\operatorname{min}}
\|\mathbf{A}(\mathbf{X})-\mathbf{d}\|_F^2   \\&+\sum_{t=1}^{N_t}{\lambda}_{g_t}R_{g}(\mathbf{z}_{g_t}), 
\text{ } s.t.\text{ } \mathbf{X} = N_{\boldsymbol{\theta}_N}(\{D_{\boldsymbol{\theta}}(\mathbf{z}_{g_t},\mathbf{z}_{c_t})\}). \label{recon_formula}
\end{split}
\end{equation}
The first term in Eq.~\eqref{recon_formula} enforces data consistency. $\mathbf{A}$ is the encoding operator, $\mathbf{X} \in \mathbb{C}^{N \times N_t}$ is the unknown image with $N$ voxels and $N_t$ "time" points, and $\mathbf{d}$ represents the acquired data. The second term regularizes the geometry latents $\mathbf{z}_{g_t}$ with $R_g$, realized via a latent diffusion prior (see Section III.A). Although both geometry and contrast priors can be learned and applied, only the geometry latent diffusion prior was adopted, due to a distribution mismatch between the training and experimental data (see Section \ref{training}).
Finally, $\mathbf{X}$ is constrained within the range space of the adaptive representation 
$N_{\boldsymbol{\theta}_N}(\{D_{\boldsymbol{\theta}}(\mathbf{z}_{g_t},\mathbf{z}_{c_t})\})$, where a refinement network $N_{\boldsymbol{\theta}_N}(.)$ further adapts decoder outputs $\{D_{\boldsymbol{\theta}}(\mathbf{z}_{g_t},\mathbf{z}_{c_t})\}$ (a set of images across "time") to the acquired data domain to reduce any representation bias. A zero-shot learning strategy is employed to estimate the parameters $\boldsymbol{\theta}_N$ using only the acquired data. 

This formulation enables effective exploitation of redundancy in multidimensional images. For image sequences with varying contrasts, the anatomy remains unchanged. We can enforce shared geometry latents, e.g., $\mathbf{z}_{g_1}=\mathbf{z}_{g_2}=\dots=\mathbf{z}_{g_{N_t}}$, reducing the number of unknowns (see specific cases in Section IV). Moreover, the disentangled representation allows geometry priors learned from large public datasets with many subjects but limited contrasts to be transferred to specific applications which may have richer contrast variations.

To solve Eq.\eqref{recon_formula}, we adapted the algorithm from \cite{song2023solving} to incorporate data consistency constraints during the reverse latent diffusion steps that enforces pre-learned feature-level prior. The overall algorithm is summarized in Algorithm~\ref{alg:resample}. More specifically, starting from Gaussian noise, the geometry latents $\{\mathbf{z}_{g_t,t_d+1}\}$ (at diffusion step $t_d+1$) are progressively denoised via an unconditional reverse process of a Denoising Diffusion Implicit Model (DDIM) \cite{song2020denoising}, where Tweedie’s formula provides an estimate of the clean latent $\{\bar{\mathbf{z}}_{g_t,0}\}$ at each step. At selected reverse diffusion steps, a conditional update was applied to adjust $\{\mathbf{z}_{g_t}\}$, $\{\mathbf{z}_{c_t}\}$ and $\boldsymbol{\theta}_N$ by minimizing the data consistency loss as follows: 
\begin{equation}
\begin{split}
&\{\hat{\mathbf{z}}_{g_t}\},\{\hat{\mathbf{z}}_{c_t}\},\hat{\boldsymbol{\theta}}_N = \\&
\arg \underset{\{\mathbf{z}_{g_t}\},\{\mathbf{z}_{c_t}\},\boldsymbol{\theta}_N}{\operatorname{min}}
\|\mathbf{A}(N_{\boldsymbol{\theta}_N}(\{D_{\boldsymbol{\theta}}(\mathbf{z}_{g_t},\mathbf{z}_{c_t})\}))-\mathbf{d}\|_F^2,
\label{recon_alg}
\end{split}
\end{equation}
which can be optimized using gradient descent. The updated geometry latents $\{\hat{\mathbf{z}}_{g_t}\}$ are then stochastically resampled back to continue the reverse diffusion process. Finally, the decoder and refinement network are applied to $\{\hat{\mathbf{z}}_{g_t}\}$ and $\{\hat{\mathbf{z}}_{c_t}\}$ to produce the reconstructed image.

\begin{figure*}[!t]
\centerline{\includegraphics[width=1\textwidth]{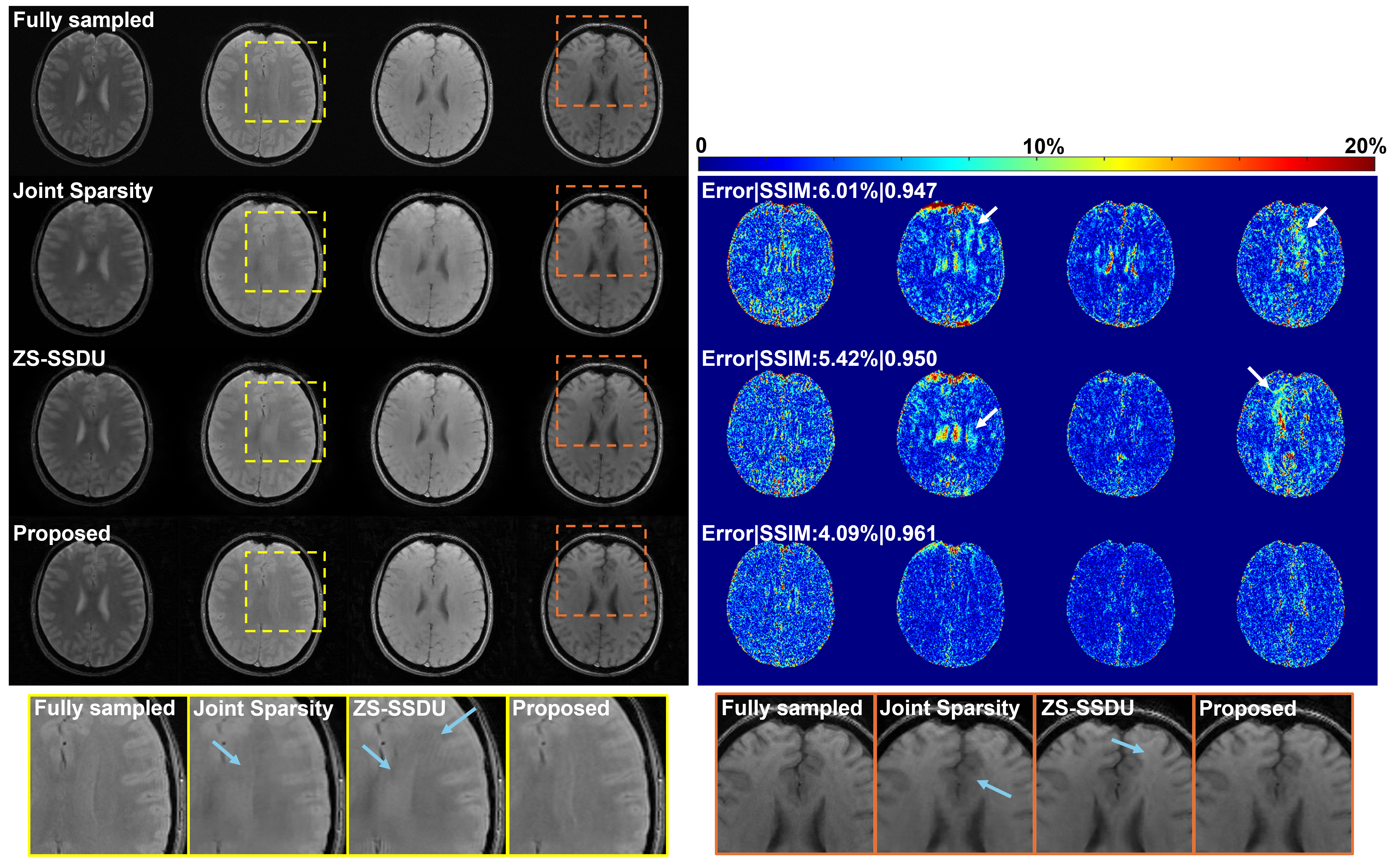}}
\caption{Reconstructed images (top left panel) from experimental $\text{T}_1$ mapping data and the corresponding error maps (top right panel) at AF=6. Images acquired with different flip angles are shown in different columns while results from different methods in respective rows. Zoomed-in regions from the images are shown in the bottom panel with artifacts that were reduced by our method indicated by blue arrows. Mean-squared errors and SSIM values are shown in the the error maps. The proposed method achieved the lowest error and highest SSIM, with most noise-like residuals.}
\label{T1_images}
\end{figure*}

Directly updating $\boldsymbol{\theta}_N$ may lead to overfitting due to the large number of parameters. Thus, we adopted the data partitioning strategy from the SSDU approach \cite{yaman2020self}. Data were partitioned into two disjoint subsets: $\mathbf{d}=\mathbf{d}_\Theta \cup \mathbf{d}_\Lambda$ (with matched forward operators $\mathbf{A}_\Theta$ and $\mathbf{A}_\Lambda$), leading to the following modified data consistency loss:
\begin{equation}
\begin{split}
\{\hat{\mathbf{z}}_{g_t}\},&\{\hat{\mathbf{z}}_{c_t}\},\hat{\boldsymbol{\theta}}_N = \\& 
\arg \underset{\{\mathbf{z}_{g_t}\},\{\mathbf{z}_{c_t}\},\boldsymbol{\theta}_N}{\operatorname{min}}
\|\mathbf{A}_\Theta(\{D_{\boldsymbol{\theta}}(\mathbf{z}_{g_t},\mathbf{z}_{c_t})\})-\mathbf{d}_\Theta\|_F^2
\\&+
\|\mathbf{A}_\Lambda(N_{\boldsymbol{\theta}_N}(\{D_{\boldsymbol{\theta}}(\mathbf{z}_{g_t},\mathbf{z}_{c_t})\},\mathbf{d}_\Theta,\mathbf{A}_\Theta))-\mathbf{d}_\Lambda\|_F^2.
\label{recon_alg_adapt}
\end{split}
\end{equation}
In this formulation, the first term enforces data consistency with respect to (w.r.t.) $\{\mathbf{z}_{g_t}\}$ and $\{\mathbf{z}_{c_t}\}$ using $\mathbf{d}_\Theta$ as the target. The second term corresponds to the SSDU reconstruction loss, where the refinement network takes both the decoder output $\{D_{\boldsymbol{\theta}}(\mathbf{z}_{g_t},\mathbf{z}_{c_t})\}$ and $\mathbf{d}_\Theta$ as input and uses the other partition $\mathbf{d}_\Lambda$ as the target. The loss is minimized w.r.t. $\{\mathbf{z}_{g_t}\}$, $\{\mathbf{z}_{c_t}\}$, and $\boldsymbol{\theta}_N$. This ``joint'' update of $\{\mathbf{z}_{g_t}\}$, $\{\mathbf{z}_{c_t}\}$, and $\boldsymbol{\theta}_N$ is computationally more efficient than solving latent first followed by the SSDU-based adaptation. The first term can be viewed as a regularization on $\{\mathbf{z}_{g_t}\}$ and $\{\mathbf{z}_{c_t}\}$ for the SSDU loss, facilitating convergence with a small number of update steps. During each gradient descent step, different partitions were created as 'data augmentation' \cite{yaman2022zero}. The refinement network $N_{\boldsymbol{\theta}_N}(\cdot)$ was implemented as an unrolled network adapted from ~\cite{MoDL}. Combining the proposed representation with the refinement network achieved better performance compared to SSDU-based reconstruction alone or reconstruction using only the learned representation (see Results).

\subsection{Training and Other Implementation Details} \label{training}
We pre-trained our representation model using the HCP and Kirby21 datasets \cite{van2013wu,landman2011multi}. $\text{T}_1$-weighted ($\text{T}_1$w) and $\text{T}_2$-weighted ($\text{T}_2$w) images from the HCP database, and $\text{T}_1$w, FLAIR, $\text{T}_1$ mapping (flip angles = $15^{\circ}$ and $60^{\circ}$), and $\text{T}_2$ mapping (TE = 30~ms and 80~ms) data from the Kirby21 database were used. The proposed reconstruction method was evaluated using in-house $\text{T}_1$ and $\text{T}_2$ mapping data, where data from one subject were used for hyperparameter tuning (validation) and data from two subjects were used for testing. Network training and reconstruction were performed on a Linux server with an NVIDIA A40 GPU and implemented in PyTorch 2.0.1. The Adam optimizer was used with a batch size of 64 for both the autoencoder and diffusion model training. Reconstruction methods based on joint sparsity and low-rank plus joint sparsity for comparison were implemented in MATLAB R2020b.

\section{Experiments and Results} \label{Results}
\subsection{Effective Feature Disentanglement}
We first evaluated the feature disentanglement of the proposed representation. As shown in Fig.~\ref{disentangle_results}, with the geometry latents of two images with the same anatomy fixed, interpolating between their respective contrast latents generated images with varying contrasts but identical anatomy, demonstrating effective separation of geometric and contrast features. These results also support that images with different contrast weightings can be produced by combining a shared geometry latent with varying contrast latents, a strong constraint for multicontrast image reconstruction. 

This disentanglement property directly motivates the application of our learned representation to MR parameter mapping, where sequences of images with varying contrasts are reconstructed and quantified. In this work, we evaluated the utility of the proposed representation and image reconstruction using quantitative $\text{T}_1$ and 
$\text{T}_2$ mapping tasks. 

\begin{figure}[!t]
\centerline{\includegraphics[width=\columnwidth]{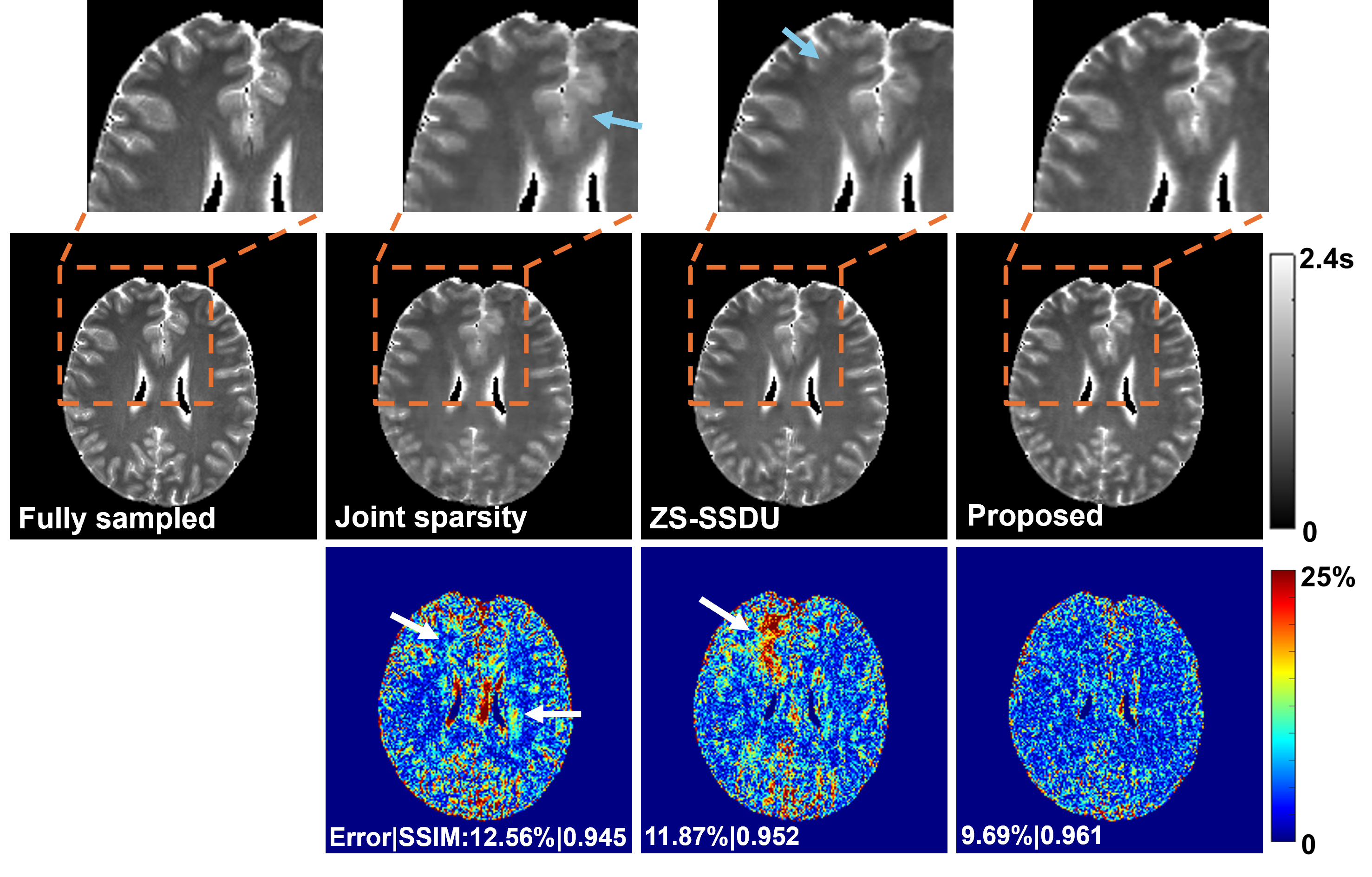}}
\caption{$\text{T}_1$ mapping results at $6\times$ acceleration (AF=6; brain masked). The proposed method preserved spatial details better with the highest mapping accuracy, as shown in the estimated $\text{T}_1$ maps (middle), zoomed-in regions (top), and the corresponding error maps (bottom). The overall errors and SSIM values for $\text{T}_1$ maps are shown.}
\label{T1_map}
\end{figure}

\par $\text{   }$

\subsection{In Vivo Experiments}
All in vivo experiments were performed with local IRB approval. $\text{T}_1$ mapping data were acquired using a variable flip angles (vFAs) protocol with a spoiled gradient echo sequence on a 3T scanner equipped with a 20-channel head coil \cite{bian2024improving}. Acquisition parameters were: TR/TE $=40 / 12 \mathrm{~ms}$, flip angles $=5^{\circ}, 10^{\circ}, 15^{\circ}, 20^{\circ}, 30^{\circ}, 40^{\circ}$, field of view $(\mathrm{FOV})=210 \times 210 \times 107 \mathrm{~mm}^3$  , and matrix size $=192 \times 192 \times 32$. 

$\text{T}_2$ mapping data were acquired using a multi-slice multi-spin-echo sequence on a 3T scanner with a 20-channel head coil \cite{peng2016accelerated}. Parameters were: TR $=3110 \mathrm{~ms}, 15$ echoes with  $\mathrm{TE}_1=11.5 \mathrm{~ms}$ and echo spacing $\Delta \mathrm{TE}=11.5 \mathrm{~ms}$. $8$ slices were acquired, each with $ \mathrm{FOV}=220 \times 200 \mathrm{~mm}^2$, matrix size $=224 \times 192$, and slice thickness $=3 \mathrm{~mm}$.

For both types of acquisitions, fully sampled data were acquired and retrospectively undersampled using a variable density random phase encoding mask (center fully sampled with Gaussian density decay toward the periphery) with different acceleration factors (AFs). Coil sensitivity maps were estimated via ESPIRiT \cite{uecker2014espirit}, using 18 central $k$-space lines from the first flip angle/TE. In the $\text{T}_2$ mapping experiment, 6 central $k$-space lines at all TEs were additionally acquired for subspace estimation \cite{zhao2015accelerated}. For accelerated $\text{T}_1$ mapping, 6 central $k$-space lines were acquired at remaining flip angles, but no subspace estimation was performed because only six flip angles were available.

\subsection{$\text{T}_1$ mapping Results}
For $\text{T}_1$ mapping, we compared the proposed method with a joint sparsity constrained reconstruction \cite{majumdar2011accelerating} and a state-of-the-art zero-shot self-supervised reconstruction method (ZS-SSDU) \cite{yaman2022zero}. We selected ZS-SSDU as a primary baseline because our method also follows a zero-shot learning paradigm, requiring no supervised re-training or fine-tuning on domain-specific datasets. Representative reconstructed images at AF=6 are shown in Fig.~\ref{T1_images}. The proposed method consistently produced reconstructions with improved quality and reduced errors across different contrasts (FAs) compared to other methods. The corresponding $\text{T}_1$ maps are shown in Fig.~\ref{T1_map}. The proposed method achieved the most accurate $\text{T}_1$ estimates w.r.t. the fully sampled reference as shown by the lowest relative $\ell_2$ errors and highest SSIM values, effectively suppressing artifacts. The zoomed-in regions further demonstrate a better preservation of structural details by the proposed method. The ZS-SSDU method, although produced lower errors than the joint sparsity method, seemed to result in strong bias in certain localized areas (Fig.~\ref{T1_map}, last row). These results demonstrated that the proposed representation can work synergistically with zero-shot self-supervised learning, leading to improved reconstruction and quantification performance. More quantitative comparisons under different AFs are shown in Fig.~\ref{T1T2error}.

\subsection{$\text{T}_2$ mapping Results}
\begin{figure}[!t]
\centerline{\includegraphics[width=\columnwidth]{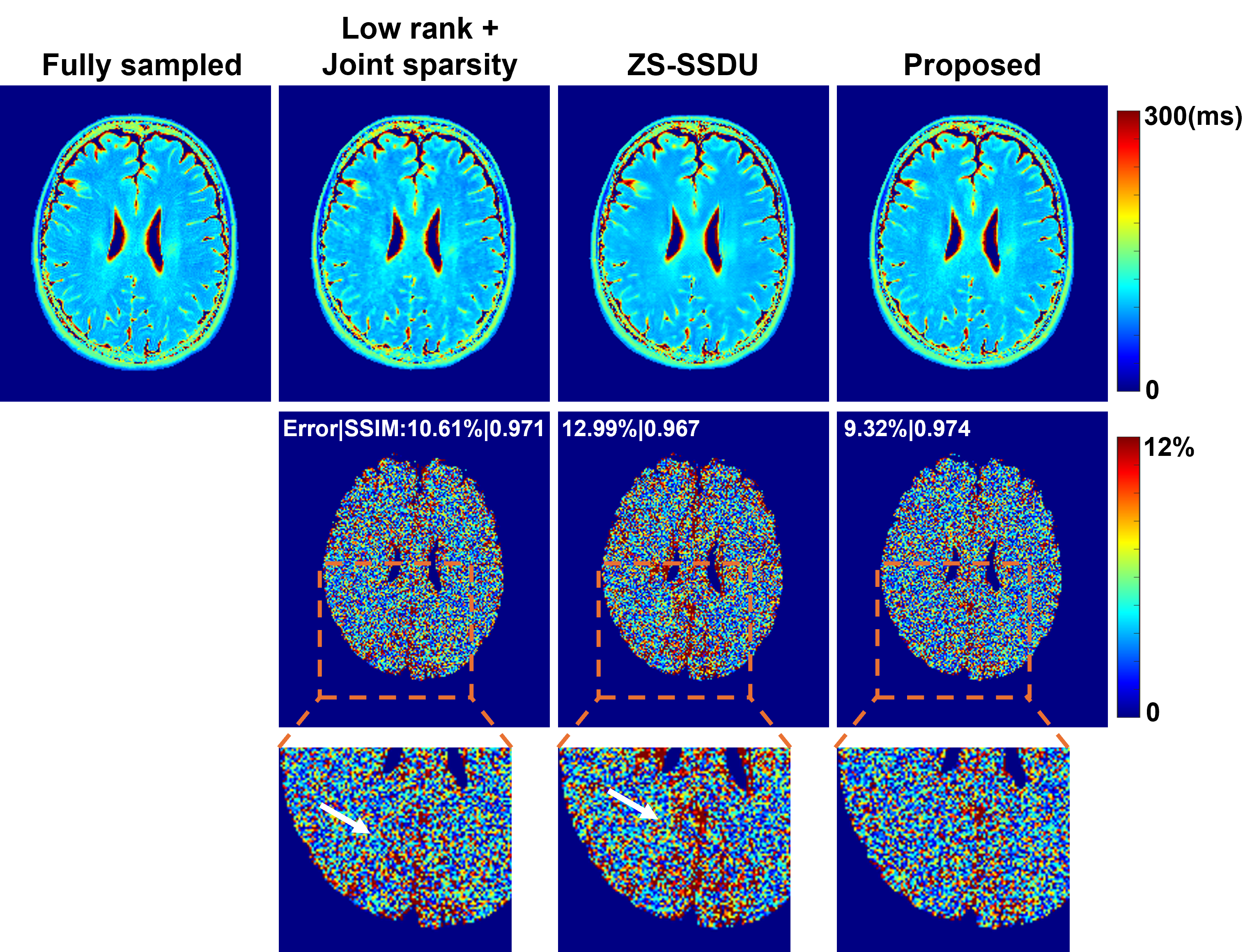}}
\caption{$\text{T}_2$ mapping results (first row) and error maps (second row) from different methods at AF=8; The overall errors and SSIM for $\text{T}_2$ maps were also shown in the error maps (top left corners). As can be seen, lower reconstruction error and better preservation of details were observed for the proposed method (last column). Zoomed-in regions further illustrate the reduced errors achieved.}
\label{T2_map}
\end{figure}
For $\text{T}_2$ mapping, the proposed method was further combined with low-rank/subspace modeling \cite{liang2007spatiotemporal}. Specifically, the underlying image $\mathbf{X}\in \mathbb{C}^{N \times N_t}$ can be modeled as: $\mathbf{X} = \mathbf{U}\hat{\mathbf{V}}$, where $\mathbf{U}\in\mathbb{C}^{N \times r}$ represents the spatial coefficients, $\hat{\mathbf{V}}\in\mathbb{C}^{r \times N_t}$ a set of temporal basis, and $r$ the subspace order/rank. $\hat{\mathbf{V}}$ can be pre-determined from central $k$-space data. This introduced a complementary constraint to the learned representation. As a result, the loss function in Eq.~\eqref{recon_alg_adapt} is changed into:
\begin{equation}
\begin{split}
&\{\hat{\mathbf{z}}_{g_t}\},\{\hat{\mathbf{z}}_{c_t}\},\hat{\boldsymbol{\theta}}_N = \\& 
\arg \underset{\{\mathbf{z}_{g_t}\},\{\mathbf{z}_{c_t}\},\boldsymbol{\theta}_N}{\operatorname{min}}
\|\mathbf{A}_\Theta(\{D_{\boldsymbol{\theta}}(\mathbf{z}_{g_t},\mathbf{z}_{c_t})\}\hat{\mathbf{V}}^{\text{H}}
\hat{\mathbf{V}})-\mathbf{d}_\Theta\|_F^2+
\\&
\|\mathbf{A}_\Lambda(N_{\boldsymbol{\theta}_N}(\{D_{\boldsymbol{\theta}}(\mathbf{z}_{g_t},\mathbf{z}_{c_t})\}\hat{\mathbf{V}}^{\text{H}},\mathbf{d}_\Theta,\mathbf{A}_\Theta,\hat{\mathbf{V}})\hat{\mathbf{V}})-\mathbf{d}_\Lambda\|_F^2.
\label{recon_alg_subspace}
\end{split}
\end{equation}

\noindent In the first term, the representation is projected onto the subspace (assuming orthonormal temporal basis). 
In the second term, instead of adapting all images directly, we perform refinement only on the spatial coefficients, 
which reduces the number of trainable parameters in the refinement network and improves the efficiency of zero-shot learning \cite{zhang2024zero}. The refinement network takes both the spatial coefficients estimate $\{D_{\boldsymbol{\theta}}(\mathbf{z}_{g_t},\mathbf{z}_{c_t})\}\hat{\mathbf{V}}^{\text{H}}$ and the temporal basis $\hat{\mathbf{V}}$ as input (for the data consistency block during unrolling). The output are refined spatial coefficients, which are then recombined with $\hat{\mathbf{V}}$ as: $N_{\boldsymbol{\theta}_N}(\{D_{\boldsymbol{\theta}}(\mathbf{z}_{g_t},\mathbf{z}_{c_t})\}\hat{\mathbf{V}}^{\text{H}},\mathbf{d}_\Theta,\mathbf{A}_\Theta,\hat{\mathbf{V}})\hat{\mathbf{V}}$ to generate the full set of reconstructed images.

\begin{figure}[!t]
\centerline{\includegraphics[width=\columnwidth]{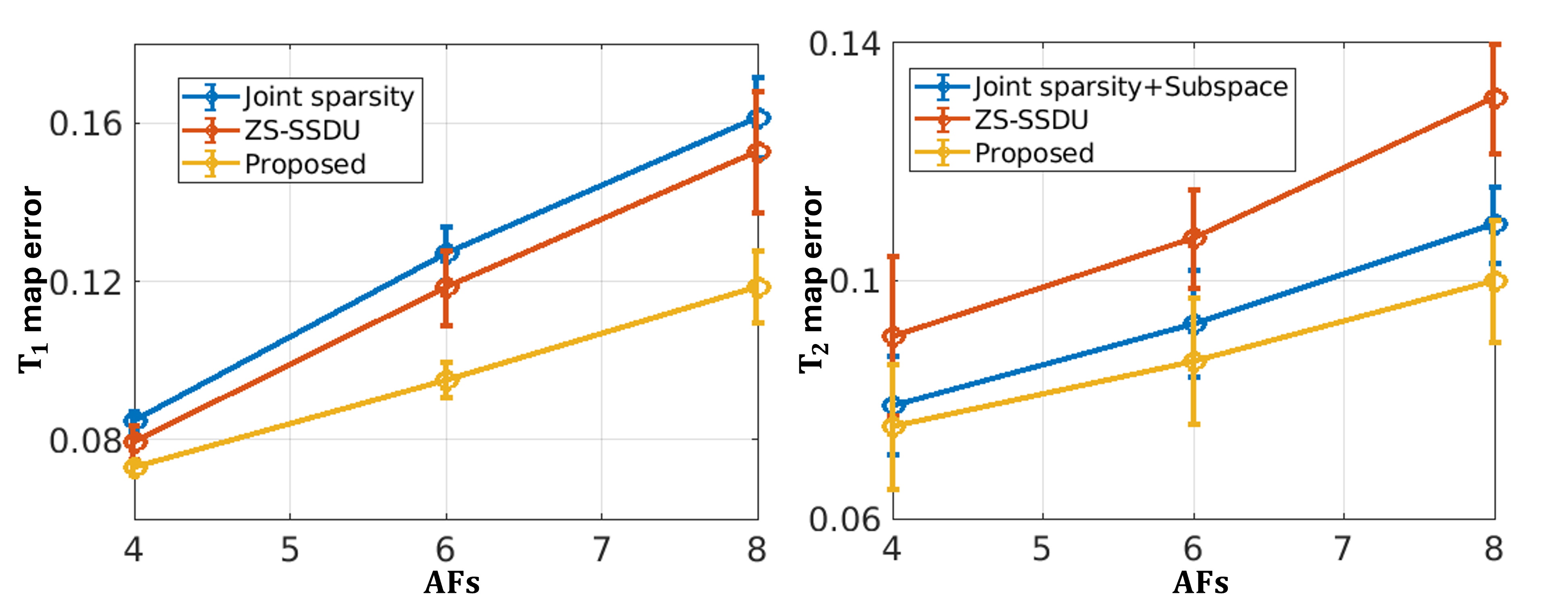}}
\caption{Parameter mapping errors at different AFs for the $\text{T}_1$ (left) and $\text{T}_2$ (right) mapping data. The proposed method consistently achieved the lowest errors across all AFs, with more pronounced improvements at higher AFs. Error bars denote standard deviation of errors across all subjects and slices.}
\label{T1T2error}
\end{figure}

Fig.~\ref{T2_map} compares $\text{T}_2$ mapping results estimated from reconstructions produced by different methods at AF = 8. The overall estimation errors and SSIM values w.r.t. the fully sampled data were computed within the brain regions with $\text{T}_2$ values lower than $500$ ms.
The proposed method integrating our disentangled representation and subspace modeling yielded superior performance over a subspace reconstruction method with joint sparsity regularization \cite{zhao2015accelerated} and the ZS-SSDU method. These results also demonstrated the flexibility of the proposed representation to be combined with complementary constraints. Fig.~\ref{T1T2error} further demonstrates lower reconstruction errors consistently achieved by the proposed method across different AFs.

\section{Discussion} \label{Discussion}
We presented an adaptive, feature-disentangled representation for multidimensional MRI reconstruction. A unique feature of our approach is its ability to more explicitly exploit feature correlations through disentanglement, where geometry and contrast variations are captured by distinct latent spaces. This design substantially reduces the degrees of freedom while enabling flexible and semantically meaningful constraints. Another key component is the second-stage refinement network, which adapts the pretrained representation to application-specific domains. By employing a zero-shot self-supervised learning strategy, the proposed refinement effectively circumvents the overfitting issue often observed in methods that jointly update both latent variables and network parameters.

\begin{figure}[!t]
\centerline{\includegraphics[width=\columnwidth]{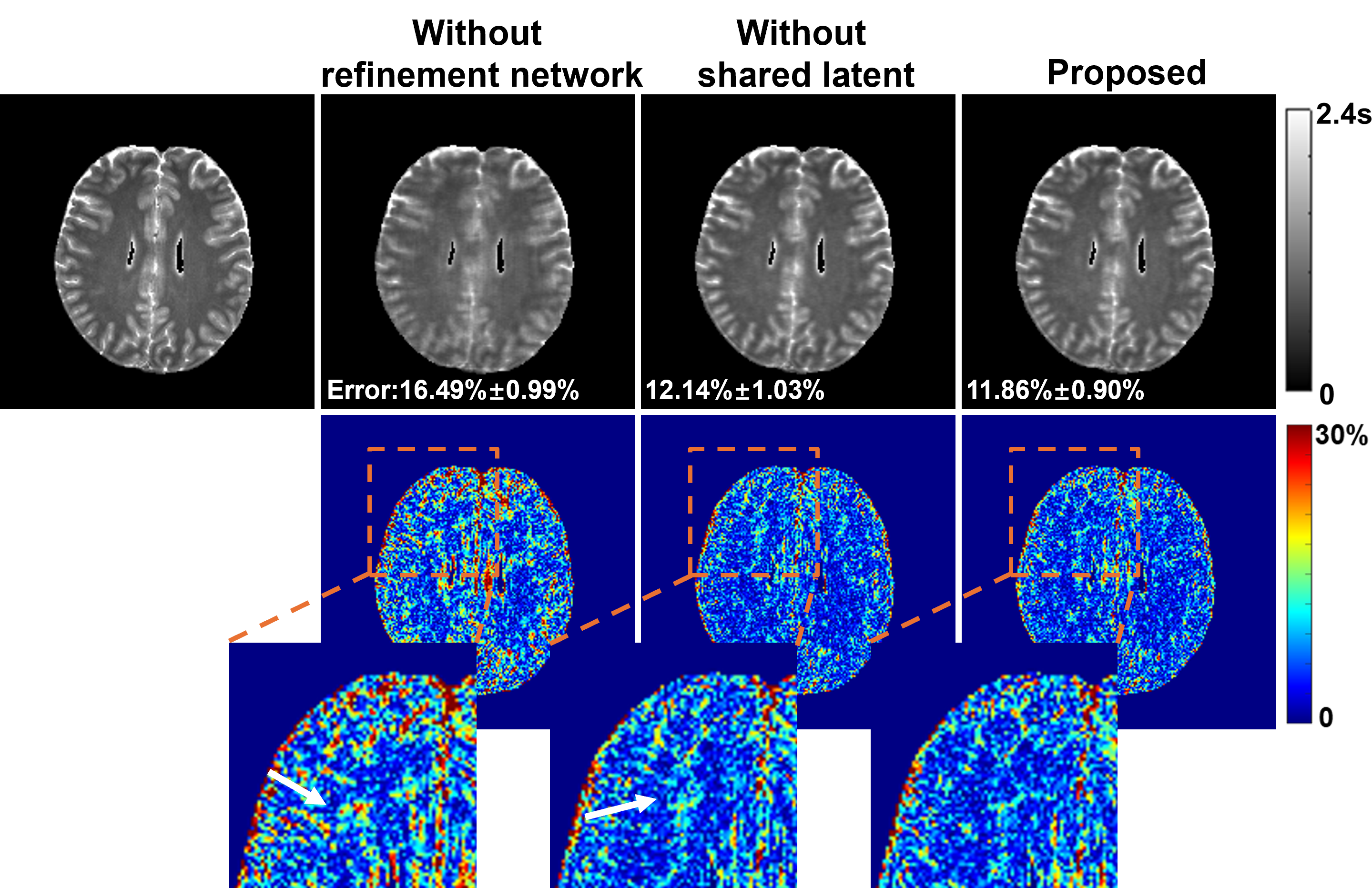}}
\caption{An ablation study using the $\text{T}_1$ mapping data (AF$=8$); Mapping errors (mean $\pm$ standard deviation) summarized across different subjects and slices are shown in the reconstructed $\text{T}_1$ maps. Reconstructions without the refinement network exhibit substantially higher errors. Compared to using independent latents, the proposed method with shared geometry latents yielded slightly lower reconstruction errors and variability. Error maps further highlight stronger artifacts when the shared latent constraint is not enforced.}
\label{ablation}
\end{figure}

One important design choice is whether geometry latents should be shared across all images with different contrasts. To evaluate this, we conducted abalation studies using $\text{T}_1$ mapping data, first comparing reconstructions using a single shared geometry latent across all FAs versus FA-dependent geometry latents. Fig.~\ref{ablation} shows that enforcing a shared geometry latent across all FAs produced better reconstruction in terms of quantitative $\text{T}_1$ maps, indicating the benefit of degrees-of-freedom reduction leveraging shared geometry features. Another ablation study shows that removing the refinement network deteriorated the results substantially (Without refinement network case in Fig.~\ref{ablation}), supporting its role in reducing representation errors and domain shifts. These results demonstrate the importance of adapting pretrained representations to application-specific domains, particularly when only limited or mismatched training data are available.

An important question that may be raised is whether geometry and contrast can be fundamentally disentangled. In practice, the definition of these two, especially for ``contrast'', is not absolute. In our formulation, the “geometry” latent essentially captures structural features shared across multiple MR contrasts, while the “contrast” latent represents image-specific variations driven by acquisition parameters or relaxation effects. While the proposed disentanglement is learned in a data-driven manner within the MRI domain, it may be extended to more general multimodal settings (e.g., MRI and CT). However, this would require rethinking how geometry and contrast are defined, as the underlying imaging physics and signal characteristics differ substantially. Achieving modality invariant geometry representations may require training on co-registered multimodal data and the design of specialized training strategies. Several recent studies have explored the disentanglement concepts for multimodal tasks such as joint MRI-CT image analysis and segmentation \cite{chartsias2019disentangled,jiang2020unified}. These methods typically aim to separate modality invariant features from modality specific appearance to facilitate down stream tasks. Extending such approaches to image reconstruction, however, would require more effective utilization of multimodal feature correlations and careful consideration of representation accuracy. Multimodal reconstruction using disentangled representations remains an open and promising direction for future research.

We adopted the StochasticResample strategy proposed in \cite{song2023solving} in our algorithm (Algorithm \ref{alg:resample}), to map the data consistency update output $\hat{\mathbf{z}}_{g_t}$ back to the diffusion sampling trajectory ${\mathbf{z}}_{g_t,t_d}$. Compared to directly adding noise to the latents (e.g., applying the forward diffusion process), this strategy  can reduce variance and lead to less noisy reconstruction \cite{song2023solving}. Mathematically, the StochasticResample can be expressed as sampling from the following distribution:
\begin{equation}
\begin{split}
&
p({\mathbf{z}}_{g_t,t_d} \mid \hat{\mathbf{z}}_{g_t},\mathbf{z}_{g_t,t_d}^{\prime})
\\& =\mathcal{N}\left(\frac{\sigma_t^2 \sqrt{\bar{\alpha}_t} \hat{\mathbf{z}}_{g_t}+\left(1-\bar{\alpha}_t\right) \mathbf{z}_{g_t,t_d}^{\prime}}{\sigma_t^2+\left(1-\bar{\alpha}_t\right)}, \frac{\sigma_t^2\left(1-\bar{\alpha}_t\right)}{\sigma_t^2+\left(1-\bar{\alpha}_t\right)} \boldsymbol{I}\right),
\label{recon_alg_resample}
\end{split}
\end{equation}
where the mean can be viewed as the a weighted combination of data consistency update output $\hat{\mathbf{z}}_{g_t}$ and the unconditional DDIM latent sample $\mathbf{z}_{g_t,t_d}^{\prime}$. $\sigma_t^2$ is defined as $\sigma_t^2=\gamma\left(\frac{1-\bar{\alpha}_{t-1}}{\bar{\alpha}_t}\right)\left(1-\frac{\bar{\alpha}_t}{\bar{\alpha}_{t-1}}\right)$. In this framework, $\gamma$ controls the tradeoff between prior consistency and data consistency.
For example, if $\gamma \rightarrow 0$, ${\mathbf{z}}_{g_t,t_d}$ will reduce to $\mathbf{z}_{g_t,t_d}^{\prime}$, which result in unconditional DDIM sampling and poor data consistency. In practice, we found a relatively large range of $
\gamma$ can result in similar reconstruction results. We set $\gamma = 35$, which worked robustly across different cases investigated here.

Another issue is reconstruction time. For the reverse process of diffusion model, we employed 500 DDIM steps and performed data consistency update every 20 steps. Each data consistency update involves 30 iterations of gradient descent to optimize $\{\mathbf{z}_{g_t}\}, \{\mathbf{z}_{c_t}\}, \boldsymbol{\theta}_N$. Under these settings, the reconstruction requires approximately 15 minutes for $\text{T}_1$ mapping and 25 minutes for $\text{T}_2$ mapping. While the proposed representation provides a strong initialization that reduces the number of steps needed for self-supervised training, and the integration with subspace modeling further decreases the number of trainable parameters, updating the parameters for $N_{\boldsymbol{\theta}_N}(\cdot)$ remains computationally expensive. Future work may address this issue by pre-training the refinement network on a small quantity of in-domain data, thereby reducing the number of online adaptation steps required.

In this work, disentanglement was enforced through an image transfer strategy. Future research may explore approaches that enforce disentanglement more explicitly in the latent space. For example, contrastive learning techniques could be employed to promote latent similarity across shared features and improve separation between geometry and contrast features \cite{radford2021learning,ouyang2021representation}. 
Alternative generative priors can be explored. For instance, models based on flow matching may accelerate the sampling process, thus accelerating reconstruction \cite{lipman2022flow}. Beyond MR parameter mapping, our method could be extended to other multidimensional MRI applications. In quantitative cardiovascular imaging, for example, modeling motion and contrast variation is essential \cite{christodoulou2018magnetic} and may require specialized network architectures with tailored feature disentanglement strategies. For applications involving inherently noisy data, such as low-field MRI or spectroscopic imaging, self-supervised denoising strategies could be integrated with autoencoder training and refinement network updates to enhance SNR.

\section{Conclusion}
This work presented a novel approach for multidimensional MRI reconstruction that leverages adaptive, feature-based disentangled representations. The learned disentangled representation effectively exploits shared and distinct image features in multidimensional data, where the associated feature priors can be pre-trained on large public data and adapted to specific applications. Our reconstruction method uses a zero-shot, self-supervised fine-tuning strategy, eliminating the need for extensive task-specific training data while maintaining strong generalization across acquisition settings. Experimental results from accelerated $\text{T}_1$ and $\text{T}_2$ mapping demonstrated the effectiveness of our proposed method. Our approach offers a promising direction for learning-based multidimensional image reconstruction with limited data for task/application-specific supervised training.
\bibliographystyle{IEEEtrans}
\bibliography{ref}

\end{document}